\pdfoutput=1

\documentclass[aps,prd,preprint,tightenlines,floatfix]{revtex4}
\usepackage{graphicx}
\usepackage{wrapfig}
\usepackage[all]{xy}
\newcommand{\nc}{\newcommand}

\nc{\beq}{\begin{equation}}
\nc{\eeq}{\end{equation}}
\nc{\bea}{\begin{eqnarray}}
\nc{\eea}{\end{eqnarray}}
\nc{\n}{\nonumber \\}

\newcommand{\acs}{\ensuremath{ \langle \sigma_{\rm a} v \rangle  }}

\usepackage[usenames]{color}
\definecolor{DarkGreen}{rgb}{0.0,0.5,0.0}

\begin{document}  

\date{March 15, 2012} 
\title{A closer look at CMB constraints on WIMP dark matter}
\author{Aravind Natarajan}
\email{anat@andrew.cmu.edu}
\affiliation{McWilliams Center for Cosmology, Carnegie Mellon University, Department of Physics, 5000 Forbes Ave., Pittsburgh PA 15213, USA}

\begin{abstract}

We use Cosmic Microwave Background data from the WMAP, SPT, BICEP, and QUaD experiments to obtain constraints on the dark matter particle mass $m_\chi$, and show that the combined data requires $m_\chi > 7.6$ GeV at the 95\% confidence level for the $\chi \chi \rightarrow b \bar b$ channel assuming $s-$wave annihilation and a thermal cross section $\langle \sigma_{\rm a} v \rangle = 3 \times 10^{-26}$ cm$^3/$s.  We examine whether the bound on $m_\chi$ is sensitive to $\sigma_8$ measurements made by galaxy cluster observations. The large uncertainty in $\sigma_8$ and the degeneracy with $\Omega_{\rm m}$ allow only small improvements in the dark matter mass bound. Increasing the number of effective neutrino-like degrees of freedom to $N_{\rm eff} = 3.85$ improves the mass bound to $m_\chi > 8.6$ GeV at 95\% confidence, for the $\chi \chi \rightarrow b \bar b$ channel. We also study models in which dark matter halos at $z<60$ reionize the Universe. We compute the Ostriker-Vishniac power resulting from partial reionization at intermediate redshifts $10<z<60$, but find the effect to be small. We discuss the importance of the large angle polarization as a complementary probe of dark matter annihilation. By performing Monte Carlo simulations, we show that future experiments that measure the $EE$ power spectrum from $20 < l < 50$ can exclude $m_\chi \sim$ 10 GeV at the 2 (3) $\sigma$ level provided the error bars are smaller than 4 (3) $\times$ cosmic variance. We show that the Planck experiment will significantly improve our knowledge of dark matter properties.

\end{abstract}

\maketitle

\section{Introduction}

The nature of dark matter remains one of the biggest puzzles in cosmology. Observational evidence from large scale structure observations, gravitational lensing, the cosmic microwave background (CMB), and galaxy rotation curves have confirmed the existence of dark matter if general relativity is the correct theory of gravitation. A well motivated dark matter candidate is the Weakly Interacting Massive Particle (WIMP), a good example of which is the Lightest Supersymmetric Particle. The dark matter direct detection experiments DAMA \cite{dama1, dama2}, CoGeNT \cite{cogent1, cogent2}, and CRESST \cite{cresst}, have obtained results that are consistent with the presence of WIMP dark matter in the Galaxy, and tentatively suggest a low dark matter mass $\sim$ 10 GeV. While these results are not without controversy, it is intriguing that WIMPs in the Galaxy may have been detected. For WIMP masses $m_\chi \sim$ 10 GeV, one may test dark matter properties by indirect detection experiments. In this article, we will restrict our discussion to the CMB, and how precise measurements of the CMB power spectra may constrain dark matter properties.

Dark matter annihilation results in energy being released in the form of standard model particles, some of which is absorbed by the surrounding gas, causing the gas to heat and ionize. The excess free electrons scatter CMB photons coming to us from the surface of last scattering. Thomson scattering of CMB photons leads to partial homogenization of temperature, resulting in a damping of the CMB power spectra on small scales. Thomson scattering also polarizes the CMB, which results in a boost in the $EE$ polarization power spectrum on large scales. 

Several authors have studied the impact of particle decay and annihilation on the CMB power spectra. Ref. \cite{pierpaoli2004} and \cite{chen_kamion2004} studied the effect of decaying dark matter on the ionization history of the Universe and the effects of a large optical depth on the CMB. Authors \cite{pad} showed that CMB polarization could detect or constrain dark matter annihilation. Ref. \cite{zhang_etal2006} and \cite{mapelli_etal2006} studied the annihilation of very light dark matter particles at early times. Ref. \cite{chuzhoy2008} examined the impact of dark matter annihilation including the clumping of dark matter, while \cite{arvi1} performed a more detailed analysis of dark matter annihilation in halos composed of light dark matter particles, and the resulting increase in optical depth. Further constraints on dark matter properties based on partial reionization and optical depth calculations were obtained by \cite{belikov_hooper2009}, and \cite{cirelli_etal2009}, while \cite{hutsi_etal2009} and \cite{yuan_etal2009} examined leptonically-annihilating dark matter models. Authors \cite{arvi2} and \cite{arvi3} studied how future large angle CMB polarization measurements could detect light dark matter, and ways in which reionization from dark matter annihilation could be distinguished from reionization from baryonic sources. Authors \cite{slatyer_etal2009} performed detailed computations of energy absorption at high redshifts. Dark matter constraints based on CMB measurements were obtained by \cite{slatyer_etal2009, kanzaki_etal2009, galli_etal2009, kim_naselsky2010}. More recently, analysis of WMAP data by \cite{hutsi_etal2011} and WMAP+ACT data by \cite{galli_etal2011} have placed stringent constraints on WIMP dark matter with mass $m_\chi < 10$ GeV. Authors \cite{finkbeiner_etal2011} performed a principal component analysis of energy injection due to dark matter annihilation, with relevance to current and future CMB observations.

In this paper, we build upon the existing literature in many ways. In Section II, we provide an introduction to WIMP annihilation and how it influences the reionization history of the Universe, which in turn influences the CMB anisotropies. In Section III, we perform a maximum likelihood analysis using the CMB $TT$ power spectrum data from the WMAP \cite{wmap} and SPT \cite{spt} experiments, $TE$ data from WMAP, and $EE$ data from the BICEP \cite{bicep} and QUaD \cite{quad} experiments. We confirm earlier results, and show that WIMP masses $m_\chi < 7.6$ GeV may be excluded at the 95\% confidence level by the combined data, for the annihilation channel $\chi \chi \rightarrow b \bar b$. Since direct detection experiments favor WIMP masses in this range, it is important to investigate whether these bounds may be improved by combining CMB data with other experiments.  The bound on the WIMP mass is weak because the damping of the power spectra ($\sim \exp -2\tau$) may be largely offset by an increase in $A_{\rm s}$ (the amplitude of the primordial curvature power spectrum), or equivalently by increasing the amplitude of matter fluctuations ($\sigma_8$). We therefore ask whether the bounds on $\sigma_8$ obtained from galaxy cluster observations may strengthen the CMB bound on the dark matter mass $m_\chi$. Unfortunately, the large $\sigma_8 \, \Omega_{\rm m}$ degeneracy of current galaxy cluster observations does not permit a significant improvement on the dark matter mass bound. We show that varying the number of effective number of neutrino-like degrees of freedom does have an effect on the likelihood function. For the best fit value $N_{\rm eff} = 3.85$ obtained by the SPT collaboration \cite{spt}, we find that the mass bound improves to $m_\chi > 8.6$ GeV, compared to the bound $m_\chi > 7.6$ GeV for the standard case of $N_{\rm eff} = 3.04$.

In Section IV, we revisit partial ionization by dark matter annihilation at intermediate redshifts, but this time accounting for dark matter halos. We show that it is possible to substantially reionize the Universe with suitable halo parameters. We compute the Ostriker-Vishniac / linear kinetic Sunyaev-Zeldovich power spectrum due to partial ionization by dark matter annihilation at intermediate redshifts $10<z<60$. The effect is however quite small, and mostly indistinguishable from the standard scenario with no reionization at these redshifts. In Section V, we show that the $EE$ power spectrum provides information complementary to what may be obtained with the $TT$ power spectrum. We perform Monte Carlo simulations, and show that light ($\sim$ 10 GeV) dark matter may be strongly constrained by measuring the $EE$ power spectrum at multipoles $l \gtrsim 20$.

\section{Dark matter annihilation}

Let us consider the case where all of the dark matter is composed of Weakly Interacting Massive Particles (WIMPs). The lightest WIMP needs to be stable or very long lived in order to be a good dark matter candidate. However, dark matter particles can annihilate in pairs, producing standard model particles. The dark matter relic density observed today $\Omega_\chi h^2 \approx 0.1$ is obtained if the annihilation cross section $\langle \sigma_{\rm a} v \rangle \approx$ 1 picobarn$\times c$, a value often considered to be a natural cross section for a weakly interacting particle. We will assume here that this is the case. We will also assume that the annihilation cross section is $s$-wave dominated, i.e. $\acs$ is independent of temperature (or relative velocity $v$).

Let $\rho_\chi (z)$ be the dark matter density at redshift $z$. The number of dark matter particles of mass $m_\chi$ per unit volume is $n_\chi = \rho_\chi / m_\chi$. The probability of WIMP annihilation per unit time is $\acs n_\chi$, and the number of dark matter pairs per unit volume is $n_\chi / 2$. The energy released per WIMP annihilation = $2 m_\chi c^2$, and a fraction $f_{\rm em}$ of this energy consists of electromagnetic particles which interact with the surrounding gas. Thus, the useful energy released per unit volume and per unit time is given by:
\beq
\frac{dE}{dtdV} = f_{\rm em} \, \frac{\rho_\chi^2(z) \acs}{m_\chi}.
\label{wimp_ann}
\eeq
Eq. (\ref{wimp_ann}) assumes that particles and antiparticles are identical. If particles are distinct from antiparticles, one must multiply Eq. (\ref{wimp_ann}) by $1/2$. For free dark matter particles (i.e. particles not bound in halos), $\rho^2_\chi$ is given by
\beq
\rho^2_{\rm free}(z) = (1+z)^6 \rho^2_{\rm crit} \Omega^2_\chi,
\label{free}
\eeq
where $\rho_{\rm crit}$ is the critical density, and $\Omega_\chi$ is the dark matter fraction today. WIMP annihilation typically produces a number of particles, each with its own energy spectrum. Let us direct our attention to low mass WIMPs, i.e. $m_\chi \sim$ 10 GeV, in agreement with the results obtained by the DAMA, CoGeNT, and CRESST direct detection experiments. Authors \cite{lnm1, lnm2, lnm3, lnm4} have considered light neutralinos within the context of an effective MSSM theory that does not assume gaugino mass unification at the GUT scale. For light non-relativistic WIMPs, annihilation to W, Z, or Higgs bosons is forbidden by conservation of energy.  Direct annihilation to photons or gluons is one-loop suppressed if WIMPs have no electric or color charge. For the special case of Majorana type fermion dark matter, annihilation to light fermions of mass $m_f$ is suppressed by a factor $(m_f/m_\chi)^2$ due to helicity conservation. Thus, for Majorana WIMPs, particle annihilation to the most massive fermion pair is favored. For WIMP masses $m_\chi \gtrsim 5$ GeV, this channel is $\chi \chi \rightarrow b \bar b$. The dominant leptonic channel is $\chi \chi \rightarrow \tau^+ \tau^-$ and the sub-dominant hadronic channel is $\chi \chi \rightarrow c \bar c$.

Fig. 1 shows the energy spectrum of neutrinos (summed over flavors) from WIMP annihilation to $b\bar b$, $c\bar c$, and $\tau^+ \tau^-$ from \cite{spectra, ciafaloni}. The fraction of energy released in the form of neutrinos is calculated as:
\beq
f_\nu = \frac{1}{m_\chi c^2} \; \int dx \, x \, \frac{dN}{dx},
\eeq
where $x = E/m_\chi$ is the dimensionless energy, and $(dN/dx) \, \Delta x$ is the number of neutrinos with dimensionless energy in the range $[x,x+\Delta x]$. The spectral function $dN/dx$ is normalized such that $f_\nu \leq 1$. The electromagnetic fraction $f_{\rm em}$ defined in Eq. (\ref{wimp_ann}) = $1 - f_\nu$. For the 3 channels considered in Fig. 1, we find $f_{\rm em}$ = 0.74 for $\chi \chi \rightarrow b \bar b$, $f_{\rm em}$ = 0.78 for $\chi \chi \rightarrow c \bar c$, and $f_{\rm em}$ = 0.65 for $\chi \chi \rightarrow \tau^+ \tau^-$. As an extreme example, we note that WIMP annihilation directly to $e^+e^-$ would result in $f_{\rm em} \approx 1$, however, this channel is unimportant for Majorana fermions. Dirac fermion dark matter particles do not suffer from helicity suppression, and may annihilate directly to $e^+e^-$, but an extra factor of $1/2$ (since particles annihilate only with antiparticles) in Eq. (\ref{wimp_ann}) means that the bound is weaker by a factor of 2. If there is significant asymmetry between particles and antiparticles, the bound is weaker still. If the dark matter is the lightest Kaluza-Klein particle (in theories with extra dimensions), the annihilation to light fermions is not helicity suppressed.

\begin{figure}[!h]
\begin{center}
\scalebox{0.8}{\includegraphics{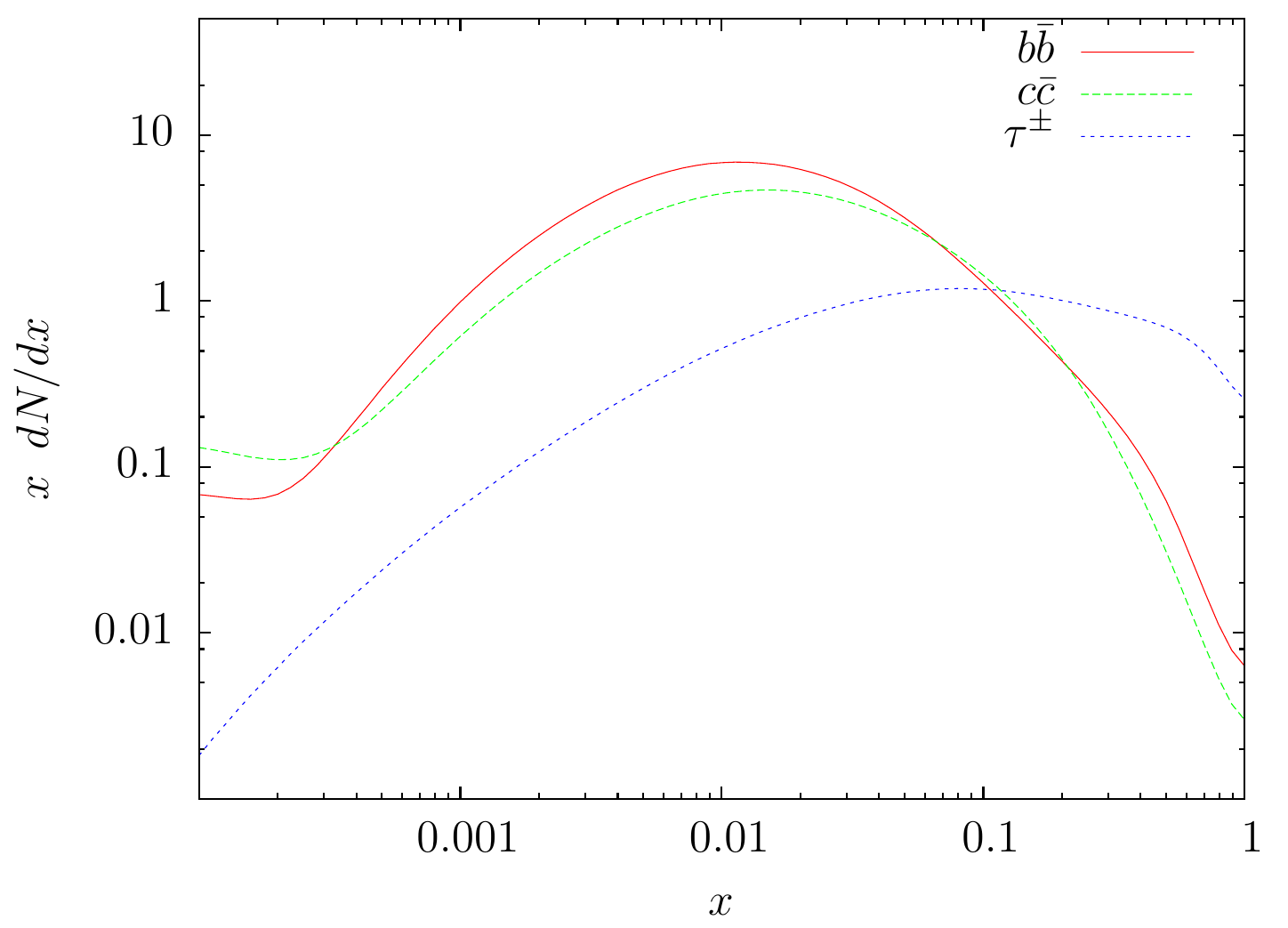}}
\end{center}
\caption{
Energy spectrum of neutrinos from WIMP annihilation to $b\bar b$, $c\bar c$, and $\tau^+ \tau^-$ channels (the spectra from $\nu_{\rm e}$, $\nu_\mu$ and $\nu_\tau$ channels are added together). $x = E/m_\chi$, and $f_{\rm em}$ = 1 - $f_\nu$ where $f_\nu$ is the neutrino fraction. For the $b\bar b$, $c\bar c$, and $\tau^+ \tau^-$ channels, we find $f_{\rm em} \approx 0.74, 0.78$, and $0.65$ respectively.
 \label{fig1} }
\end{figure}

Let us first consider dark matter annihilation at high redshifts. The mean free path for photons/charged particles scattering with background atoms = $\Lambda \sim [ n_{\rm b}(z) \sigma(E) ]^{-1}$, where $n_{\rm b}$ is the baryon number density at redshift $z$, $\sigma$ is the scattering cross section, and $E$ is the energy of the photon/charged particle  (we are assuming here that the wavelength of the radiation is much smaller than the Bohr radius, and hence neutral Hydrogen atoms may be treated as simply protons and electrons). For radiation to be effectively absorbed, it is necessary that $\Lambda$ is smaller than the Hubble horizon, i.e $\Lambda < c/H(z)$, where $H(z)$ is the Hubble parameter at redshift $z$. For $\sigma(E) \sim$ the Thomson cross section $\sigma_{\rm T}$, we find $\Lambda < c/H$ for $z>40$. Note that prompt photons from WIMP annihilation may have very high energies, and hence small cross sections, however, inverse Compton scattering with CMB photons at high redshifts is very efficient in producing a flux of lower energy photons. Thus, for redshifts $z \gg 40$, we may make the ``on the spot'' approximation, i.e. the assumption that the energy released by WIMP annihilation at a redshift $z$ is absorbed at approximately the same redshift. For redshifts $z \lesssim 40$, we must study the radiative transfer carefully to obtain the energy absorbed at a given redshift. We will discuss this case later. We note that even at high redshifts, only a fraction ${\bar f_{\rm abs}} \approx 0.85$ of the energy is absorbed \cite{slatyer_etal2009}, where the bar indicates an average over redshifts.

At high redshifts, the energy absorbed per atom per unit time, at a redshift $z$ is given by:
\bea
\xi(z) &=& {\bar f_{\rm abs}} \, f_{\rm em} \frac{\rho_\chi^2(z) \acs}{m_\chi \, n_{\rm b}(z)}  \n
&=& \frac{ {\bar f_{\rm abs}} \, f_{\rm em} \, \bar m \, \acs }{m_\chi}  \; \left ( \frac{\rho_{\rm crit}}{h^2} \right ) \, \frac{\left ( \Omega_\chi h^2 \right )^2 }{\Omega_{\rm b} h^2} \, \left(1 + z \right )^3 \n
&\approx& 3.9\times 10^{-4} \; \frac{ {\rm eV} }{ {\rm Myr} } \; \left (\frac{1+z}{100} \right )^3 \left ( \frac{10 \, {\rm GeV}}{m_\chi} \right ),
\label{xi}
\eea
for $\Omega_{\rm b} h^2$ = 0.0226, $\Omega_\chi h^2$ = 0.111, $\acs = 3 \times 10^{-26}$ cm$^3$/s, and assuming ${\bar f_{\rm abs}} = 0.85, f_{\rm em} = 0.74$.  $\bar m$ is the mean mass per atom $\approx 1.22 \, m_{\rm p}$ where $m_{\rm p}$ is the proton mass and assuming 24\% Helium. $h = H_0/(100$ km/s/Mpc) is the dimensionless Hubble parameter at the present epoch, $\Omega_\chi$ and $\Omega_{\rm b}$ are the dark matter and baryon densities today. 

Energy released by WIMP annihilation partially ionizes gas in the surrounding medium. The fraction of the absorbed energy that goes into ionization has been studied by \cite{furlanetto_2010, shull_vansteen, kanzaki_energy, valdes}. Detailed computations \cite{furlanetto_2010} indicate that a fraction $\eta_{\rm ion} \approx 0.4$ of the absorbed energy is used for ionization, for small ionization fractions $x_{\rm ion}$. A fraction $\eta_{\rm heat} \approx 0.2$ of the energy results in heating, for small $x_{\rm ion}$, with the remaining energy going into collisional excitations. The ionization fraction $x_{\rm ion}(z)$ and gas temperature $T_{\rm gas}(z)$ at a given redshift are computed by solving together, the two equations:
\bea
-(1+z) H(z) \frac{dx_{\rm ion}(z)}{dz}  &=& \mu \left [1-x_{\rm ion}(z) \right ] \eta_{\rm ion} (z)  \xi(z)   - n(z) x^2_{\rm ion}(z) \alpha(z)  \n
-(1+z) H(z) \frac{dT(z)}{dz}  &=& - 2T(z) H(z)  +  \frac{2 \eta_{\rm heat}(z)} {3 k_{\rm b}} \, \xi(z)  +   \frac{ x_{\rm ion}(z) \left [ T_\gamma(z) - T(z) \right ]}{t_{\rm c} (z)}.
\label{ion_T}
\eea
$\mu \approx 0.07$ eV$^{-1}$ is the inverse of the average ionization energy per atom, assuming 76\% H and 24\% He, neglecting double ionization of Helium \cite{arvi1}. $\alpha$ is the case-B recombination coefficient, $T_\gamma$ is the CMB temperature, and $t_{\rm c}$ is the Compton cooling time scale $\approx$ 1.44 Myr $[30/(1+z)]^4$. The last term in the temperature evolution equation accounts for the transfer of energy between free electrons and the CMB by compton scattering \cite{coupling, recfast1, recfast2}. We used $x_{\rm ion} \ll 1$ and ignored the Helium number fraction in the temperature coupling term. In practice, we compute $x_{\rm ion}$ and $T_{\rm gas}$ using a modified version of the publicly available RECFAST program \cite{recfast1, recfast2}.

\section{CMB data and constraints on the WIMP mass}

Let us first consider the impact of dark matter annihilation at high redshifts on the CMB. At redshifts $z>60$, the contribution from dark matter halos is negligible, so we consider just the smooth component. Energy from dark matter annihilation partially ionizes the surrounding gas. This leads to an excess of free electrons which scatter off CMB photons, resulting in partial homogenization of the CMB temperature. The power spectra are thus damped relative to the standard $\Lambda$CDM model by a factor $\exp(-2\tau_{\rm dm})$ where $\tau_{\rm dm}$ is the excess optical depth due to free electron-CMB photon scattering. 
\begin{figure}[!h]
\begin{center}
\scalebox{0.6}{\includegraphics{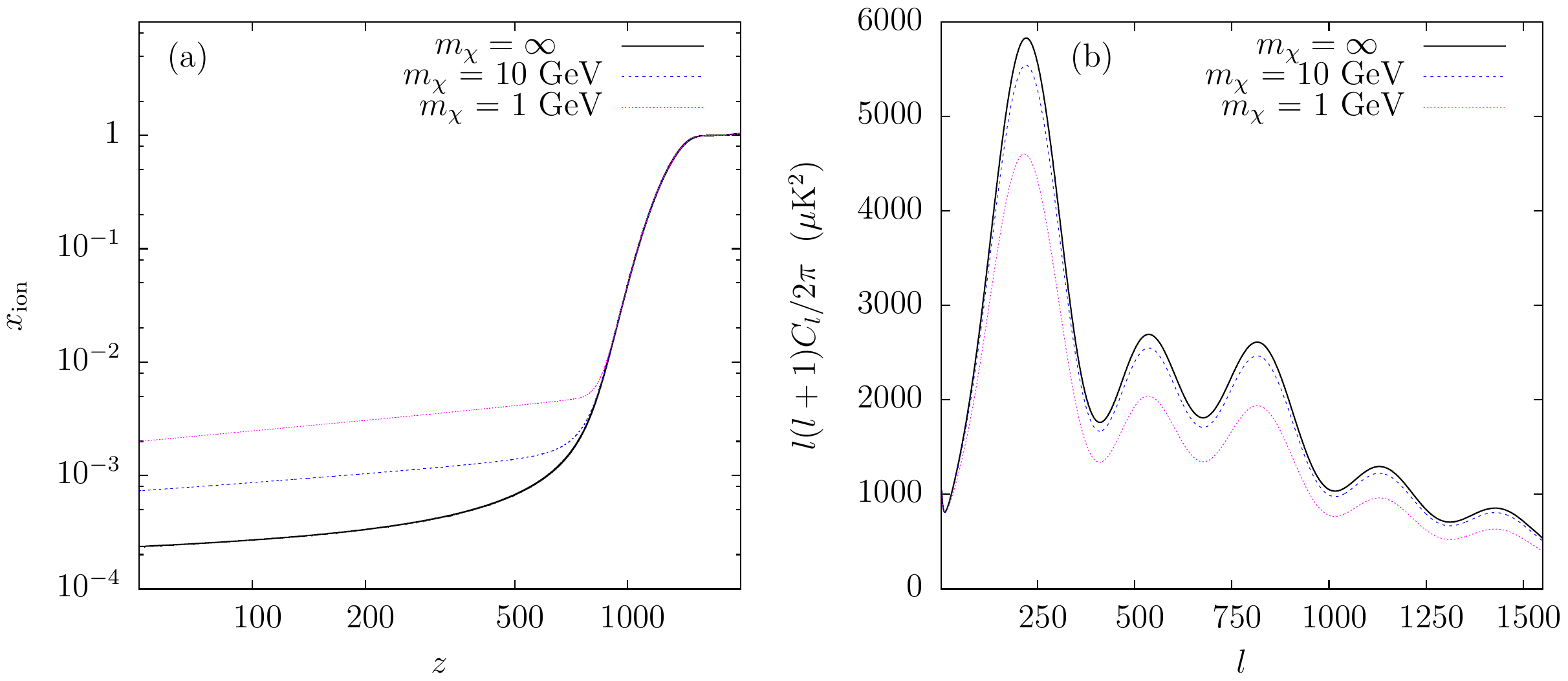}}
\end{center}
\caption{ 
The effect of WIMP annihilation on the ionized fraction and the CMB $TT$ power spectrum. (a) shows the evolution of the ionized fraction with redshift, for the standard scenario without dark matter annihilation, as well as 2 models with light WIMPs. (b) shows the $TT$ power spectrum for the 3 cases. A larger ionized fraction results in more scattering of CMB photons with free electrons, and hence lower power. The $\exp(-2\tau)$ damping may be partially offset by a larger value of $A_{\rm s}$.
 \label{fig2} }
\end{figure}
Fig. \ref{fig2}(a) shows the evolution of the ionized fraction for the standard scenario without WIMP annihilation (solid black curve), as well as models with light WIMPs. At very high redshifts $z>800$, particle annihilation does not result in excess ionization because the available ionization energy due to particle annihilation is small compared to the radiation energy. At lower redshifts, the effect of particle annihilation is clear. Fig. \ref{fig2}(b) shows the predicted $TT$ power spectrum for the model with no dark matter annihilation, as well as for the light dark matter models.  The total optical depth $\tau(z)$ up to a redshift $z$ is defined as:
\beq
\tau(z) = \int_0^z \frac{dz'}{(1+z') H(z')}  c \, \sigma_{\rm T} n_{\rm b}(z') x_{\rm ion}(z').
\label{opt_depth}
\eeq
At high redshifts, $H(z) \propto (1+z)^{3/2}$, and hence the integrand $\propto (1+z)^{1/2} \; x_{\rm ion}(z)$. We see that even small changes in $x_{\rm ion}$ at high redshifts can alter the optical depth $\tau$, resulting in a damped power spectrum.

A major difficulty in observing the predicted damping is that the power spectrum $\propto A_{\rm s} \exp(-2\tau)$, where $A_{\rm s}$ is the amplitude of the gauge invariant primordial scalar curvature power spectrum at a pivot scale $k_{\rm pivot}$ = 0.05 Mpc$^{-1}$. Unless the value of $A_{\rm s}$ is fixed by other experiments, this degeneracy implies that a damped power spectrum may be partially offset by a proportionally larger value of $A_{\rm s}$, or equivalently, by a larger $\sigma_8$. The degeneracy is not exact however, as causality requires that scales larger than the horizon at the epoch of particle annihilation be unaffected. Since the horizon $\sim (1+z)^{-3/2}$, particle annihilation at high redshifts affects the CMB on moderate and small scales, but leaves the large angle (small $l$) power spectrum unaffected. We will use this scale dependence to constrain the WIMP mass.

\begin{figure}[!h]
\begin{center}
\scalebox{0.7}{\includegraphics{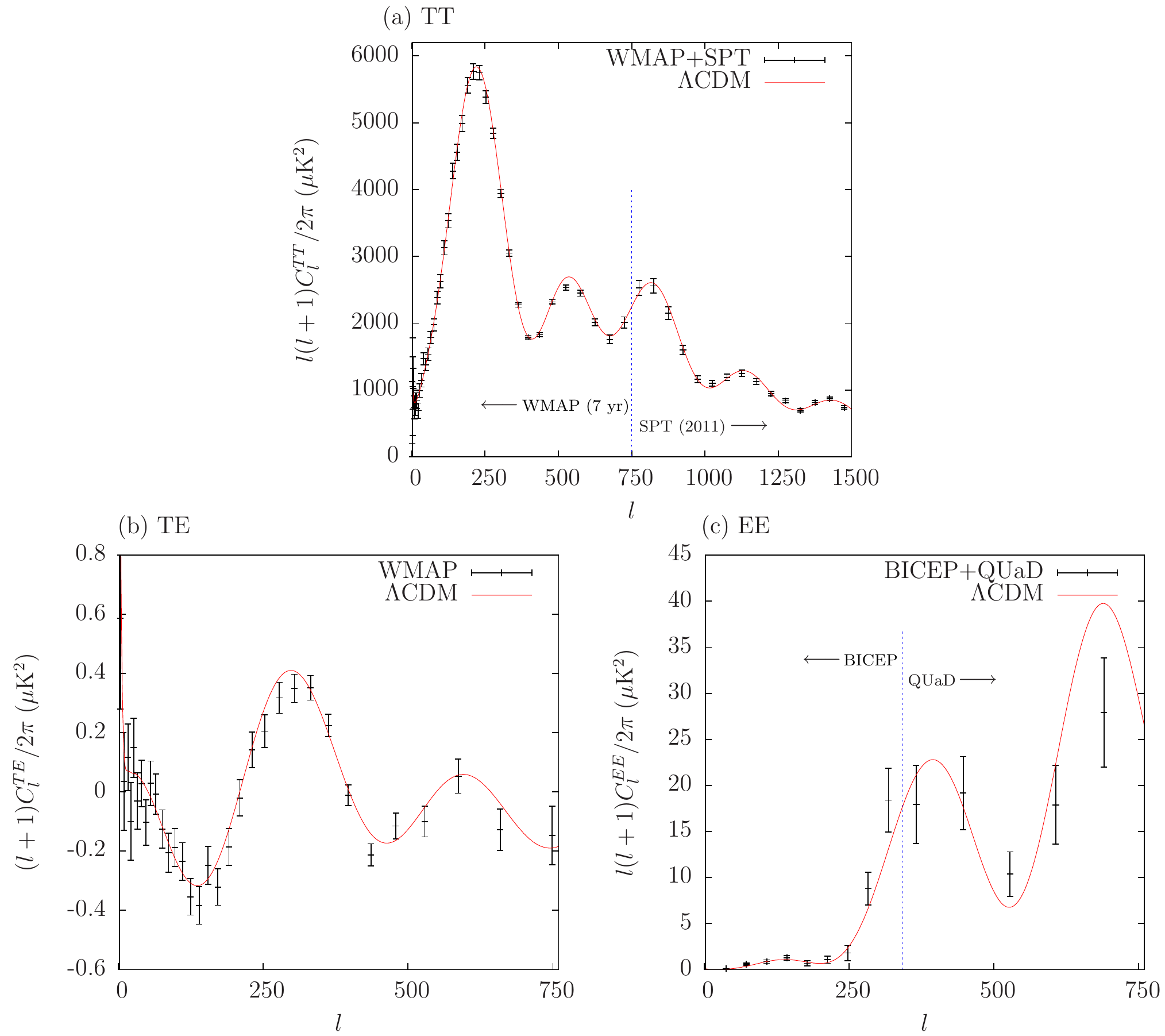}}
\end{center}
\caption{ The CMB data. (a) shows the $TT$ power spectrum. For $l<750$ we use the WMAP 7 year data release, while for $l>750$, we use data from the SPT experiment. (b) shows the $TE$ power spectrum from the WMAP 7 year data set. (c) shows the $EE$ power spectrum from the BICEP (for $l<350$) and QUaD (for $l>350$) experiments. Shown in red is the prediction of the standard $\Lambda$CDM model without WIMP annihilation included. Following the WMAP convention, we plot $(l+1)C_l/2\pi$ for the $TE$ power spectrum and $l(l+1)C_l/2\pi$ for the $TT$ and $EE$ power spectra. 
 \label{fig3} }
\end{figure}

Fig. \ref{fig3} shows the data set that we will be considering. (a) shows the $TT$ data from the publicly available 7-year data release from the WMAP mission combined with data from the SPT experiment. For $l>750$, we find that the SPT data has smaller error bars than WMAP, while for $l<750$, the WMAP data is better. We do not consider data for $l>1500$ due to complications from foreground sources, and secondary processes such as lensing. (b) shows the $TE$ power spectrum data from WMAP. (c) shows the $EE$ power spectrum from the BICEP (for $l<350$) and QUaD (for $l>350$) experiments. We use 53 data points from the $TT$ power spectrum data set, 33 points from the $TE$ data set, and 15 from the $EE$ data set. Also shown in red are the predictions for the standard $\Lambda$CDM cosmology without the effect of dark matter annihilation.

As mentioned previously, we consider the simple scenario in which all of the dark matter is made up of WIMPs, and the annihilation cross section $\acs$ is given by the thermal value, and is assumed to be independent of temperature (i.e. the annihilation is $s$-wave dominated). We obtain constraints on the WIMP mass by performing a maximum likelihood analysis using the publicly available CMB Boltzmann code CLASS \cite{class1, class2}, varying the following parameters: $\{ m_\chi, A_{\rm s}, n_{\rm s}, h, \Omega_{\rm b} h^2, \Omega_{\rm m}h^2 \}$. $n_{\rm s}$ is the scalar spectral index, $h$ is the Hubble parameter today in units of 100 km/s/Mpc. $\Omega_{\rm b}$ and $\Omega_{\rm m}$ are the baryon and matter density fractions at the present epoch. Single step reionization at $z_\ast=10.5$ is assumed. For simplicity, we have only varied the cosmological parameters that are most affected by dark matter annihilation. As discussed earlier, the parameter $A_{\rm s}$ is largely degenerate with $m_\chi$. The parameters $n_{\rm s}, \Omega_{\rm b} h^2$ and $\Omega_{\rm m} h^2$ modify the location and height of the peaks and hence may compensate for the effect of $m_\chi$. Varying $h$ affects the best fit value of $A_{\rm s}$, as well as the value of the likelihood function.  $\Omega_{\rm b} h^2$ and $\Omega_{\rm m} h^2$ also affect the ionization directly (Eq. (\ref{xi})). We set the equation of state of dark energy $\omega = -1$. We also assume zero curvature, and no running of the spectral index. The primordial Helium fraction is set to 0.24. We first assume the number of (massless) neutrino-like degrees of freedom $N_{\rm eff}$ = 3.04. We later examine the effect of a larger value of $N_{\rm eff}$.

We begin by obtaining the best fit values of $\{ A_{\rm s}, n_{\rm s}, h, \Omega_{\rm b} h^2, \Omega_{\rm m}h^2 \}$ by setting $m_\chi = \infty$ and minimizing the value of $\chi^2$:
\beq
\chi^2 = \sum_l \left ( \frac{ C_l ({\rm data}) - C_l ({\rm theory}) }{\delta C_l} \right )^2,
\eeq
where $\delta C_l$ is the $1-\sigma$ error bar in $C_l$. With the combined $TT + TE + EE$ data, we obtain a best fit value $\chi^2_{\rm min}$ = 92.5/96 d.o.f. (101 data points, 5 fitting parameters with $m_\chi$ fixed to $\infty$), with best fit values $\{ 10^9 A_s = 2.24,  n_s =  0.97, h = 0.69,  \Omega_{\rm m} h^2 = 0.1395,  \Omega_{\rm b} h^2 = 0.0225 \}$. With the $TT+TE$ data, we obtain a best fit value $\chi^2_{\rm min}$ = 75.1/77 d.o.f., while for the $TT$ data alone, we find $\chi^2_{\rm min}$ = 46.8/48 d.o.f.

\begin{figure}[!h]
\begin{center}
\scalebox{0.6}{\includegraphics{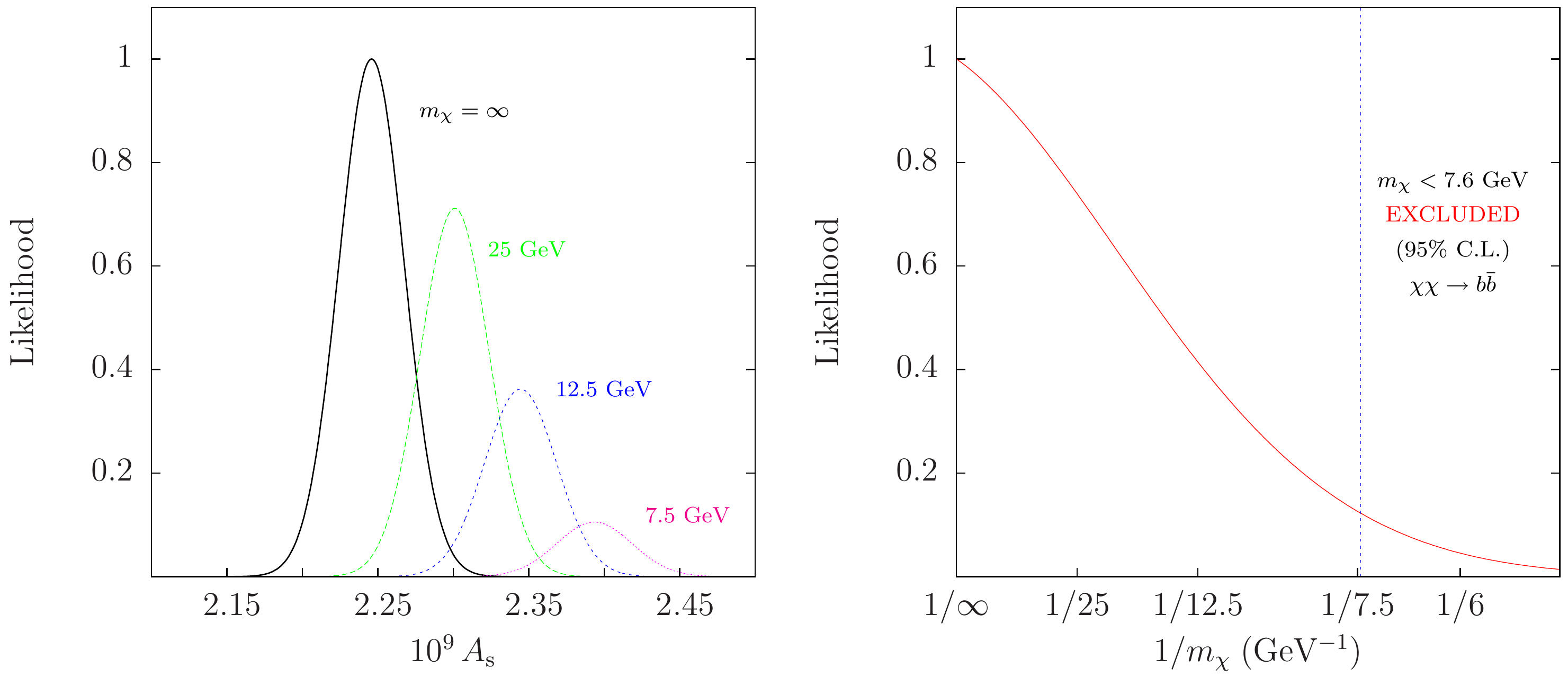}}
\end{center}
\caption{ 
The likelihood function and WIMP mass exclusion. (a) shows the normalized likelihood as a function of $10^9 A_{\rm s}$ marginalized over other cosmological parameters, for different WIMP masses. As the WIMP mass is increased, the likelihood function peaks for larger values of $A_{\rm s}$.  (b) shows the likelihood marginalized over $A_{\rm s}$, as a function of $1/m_\chi$. WIMP masses below 7.6 GeV are excluded at the 95\% level for the $b\bar b$ channel.
 \label{fig4} }
\end{figure}

The likelihood function $\mathcal L$ is defined as $-2 \ln \mathcal L = \chi^2$ + constant. We marginalize the likelihood by integrating over variables:
\beq
\mathcal L(a) =  \int db \, \mathcal L(a,b).
\eeq
Fig. \ref{fig4}(a) shows the normalized likelihood function for $h=0.69$ and marginalized over other cosmological parameters, as a function of $10^9 A_{\rm s}$. Shown are likelihood curves for different WIMP masses. Small WIMP masses result in a significant damping of the power spectrum, and hence require a larger $A_{\rm s}$ to compensate, resulting in curves peaking at larger values of $A_{\rm s}$. The area under each likelihood curve is a measure of how well that model fits the data. Fig. \ref{fig4}(b) shows the likelihood function marginalized over $\{ A_{\rm s}, n_{\rm s}, h, \Omega_{\rm b} h^2, \Omega_{\rm m} h^2  \}$, as a function of $1/m_\chi$. At the 95\% confidence level, we are able to exclude a WIMP mass $m_\chi < 7.6$ GeV for the specific channel $\chi \chi \rightarrow b\bar b$, assuming a thermal annihilation cross section and $s$-wave annihilation, and assuming no prior knowledge of cosmological parameters. WIMP exclusion limits for other annihilation channels may also be obtained from the above result, since $\xi \propto f_{\rm em} / m_\chi$  (see Eq. (\ref{xi})). Thus, for the $\chi \chi \rightarrow \tau^+ \tau^-$ channel, we have $f_{\rm em} = 0.65$, which excludes WIMP masses $m_\chi < 6.7$ GeV. For the extreme case of $f_{\rm em} \approx 1$ obtained for $\chi \chi \rightarrow e^+e^-$, we exclude WIMP masses $m_\chi < 10.3$ GeV. We note that ignoring the $EE$ data resulted in slightly stronger bounds on the WIMP mass, and emphasize that one must include all the available data when computing dark matter constraints. 
 
We have seen in the previous section that WIMP annihilation leads to a damping in the small scale $TT$ and $TE$ power spectra, requiring a larger value of $A_{\rm s}$ to compensate for the damping. Since $A_{\rm s} \propto \sigma^2_8$ for fixed $\Omega_{\rm m}h^2$, $\Omega_{\rm b}h^2$, and $n_{\rm s}$, this implies a proportionally large value of $\sigma_8$, where $\sigma_8$ is the RMS matter fluctuation averaged over a scale of 8 Mpc/$h$. A large matter power leads to an abundance of galaxy clusters which may be observed by means of X-ray telescopes and by CMB experiments through the thermal Sunyaev-Zeldovich effect \cite{zeldovich_sunyaev}.

The number of galaxy clusters $N(M_1, M_2)$ between masses $M_1$ and $M_2$ is given by the formula
\beq
N(M_1, M_2) = \int_{M_1}^{M_2} dM \; \int_{z_{\rm min}}^{z_{\rm max}(M)} dz \; \frac{dN}{dMdV} (z,M) \; \frac{dV}{dz} (z).
\eeq
$z_{\rm min}$ is the lowest redshift probed by the survey, and $z_{\rm max} (M)$ is determined by the minimum flux (or integrated Compton-Y parameter) required in order to make a detection of the cluster. $dN/dM dV$ is the number of clusters (or halos) per unit mass per unit volume and is obtained numerically \cite{press, st1,st2,jenkins,tinker}. This function is sensitive to cosmological parameters. $dV/dz$ is the cosmological volume-redshift  relation. Matching the predicted number of clusters with the observations yields a bound on $\sigma_8$ which is largely degenerate with $\Omega_{\rm m}$. If $\sigma_8$ is accurately determined, one may use this bound to limit the variation of $A_{\rm s}$ and hence obtain stronger bounds on $m_\chi$ using CMB observations. We convert $A_{\rm s}$ to $\sigma_8$ using the approximate fitting formula given by \cite{hujain}:
\beq
\sigma_8 = f \times \sqrt{ \frac{10^9 A_{\rm s}}{3.125} }  \left( \frac{\Omega_{\rm b} h^2}{0.024} \right )^{-1/3}  \left ( \frac{\Omega_{\rm m} h^2 }{0.14} \right )^{0.563} \left( 3.123h \right )^{(n_{\rm s}-1)/2} \left( \frac{h}{0.72} \right )^{0.693} \frac{G_0}{0.76}.
\eeq
$G_0 \approx 0.76$ is the growth factor today, and $f$ is a ``fudge factor''. Using the cosmological parameters determined by \cite{spt}, one obtains agreement between $\sigma_8$ and $A_{\rm s}$, for $f = 0.96$.

\begin{figure}[!h]
\begin{center}
\scalebox{0.6}{\includegraphics{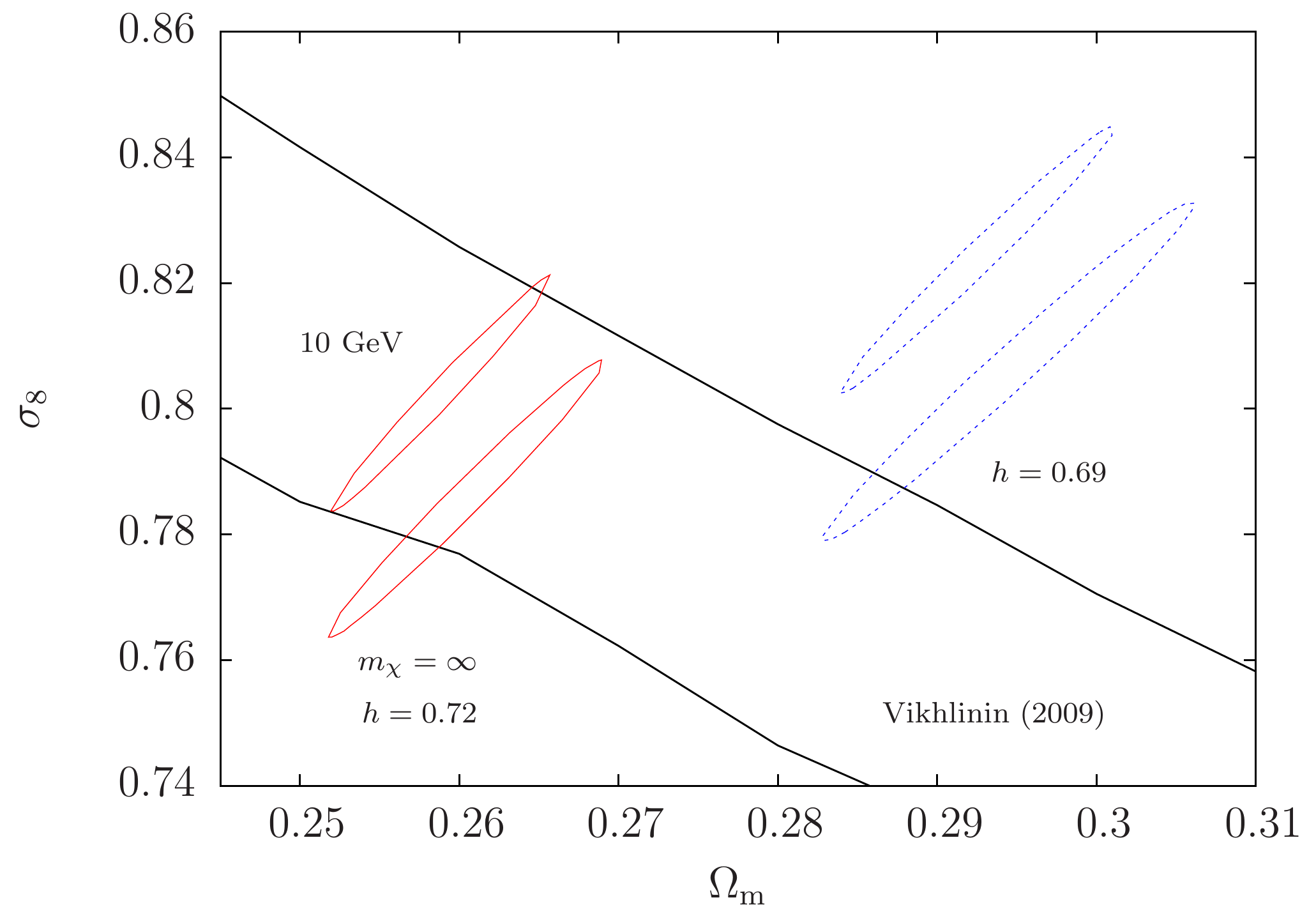}}
\end{center}
\caption{ $\Omega_{\rm m} \; \sigma_8$ contours from CMB data, along with the allowed parameter space from galaxy cluster observations by Vikhlinin et al. (region within the solid black lines). The solid (red) contours  assume $h=0.72$, while the dotted (blue) contours are for $h=0.69$. The cluster results from Vikhlinin et al. are marginalized over $h$ assuming a Gaussian prior $h=0.72 \pm 0.08$.
 \label{fig5} }
\end{figure}

Fig. \ref{fig5} shows the $\Omega_{\rm m} \,  \sigma_8$ contours for $m_\chi = \infty$ and $m_\chi$ = 10 GeV, marginalized over $A_{\rm s}, \Omega_{\rm m} h^2$, and $\Omega_{\rm b} h^2$. The solid (red) contours are plotted for $h=0.72$, while the dotted (blue) contours are plotted for $h=0.69$. Also shown are the constraints obtained by \cite{vikh1, vikh2} using \textit{Chandra} observations of X-ray galaxy clusters (the galaxy cluster contours are marginalized over $h$ using a Gaussian prior $h = 0.72 \pm 0.08$). We note that other authors, e.g. \cite{campanelli_etal} find smaller $\sigma_8$ and larger $\Omega_{\rm m}$ values by accounting for dynamical dark energy. From Fig. \ref{fig5}, we note that the uncertainty in $\sigma_8$ from galaxy cluster observations is quite large due to the degeneracy with $\Omega_{\rm m}$. Since this uncertainty is typically larger than the uncertainty in $\sigma_8$ due to dark matter annihilation, only modest improvements in the dark matter mass bound are possible with current cluster constraints.

Up to now, we have assumed $N_{\rm eff} = 3.04$ which is predicted by the standard model of particle physics. Surprisingly, both the SPT \cite{spt} and ACT \cite{act} experiments have reported deviations from the standard value, using precision measurements of CMB data on small scales. The SPT collaboration has obtained a value of $N_{\rm eff} = 3.85 \pm 0.62$, while the ACT collaboration finds $N_{\rm eff} = 5.3 \pm 1.3$.
\begin{figure}[!h]
\begin{center}
\scalebox{0.55}{\includegraphics{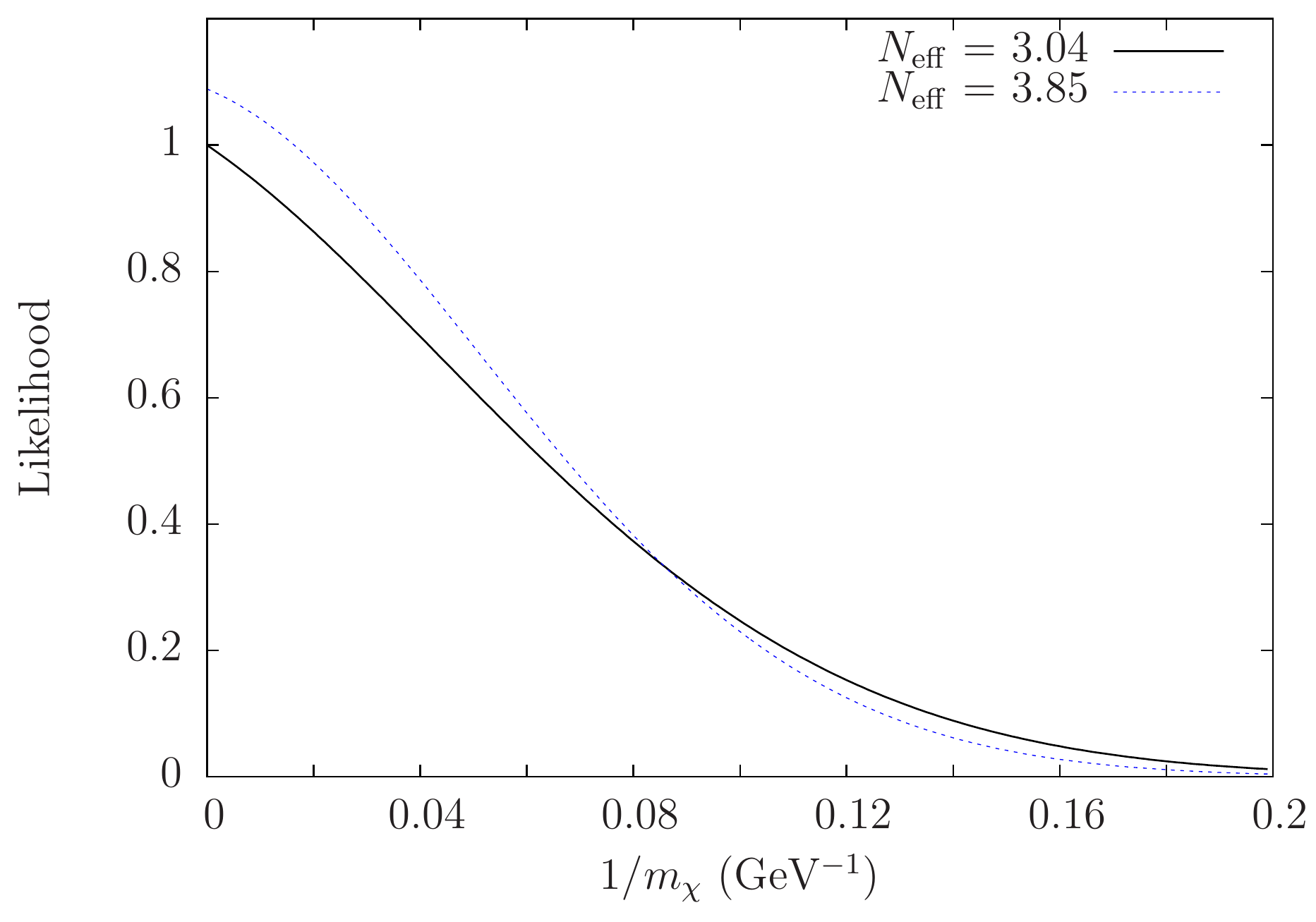}}
\end{center}
\caption{ 
The likelihood function marginalized over cosmological parameters, for $N_{\rm eff}$ = 3.04 (solid, black) and $N_{\rm eff}$ = 3.85 (dotted, blue). The $N_{\rm eff}=3.85$ case is sightly preferred over the standard model, and falls off more rapidly with decrease in $m_\chi$.  The mass bound for the $N_{\rm eff} = 3.85$ scenario is $m_\chi > 8.6$ GeV compared to the bound $m_\chi > 7.6$ GeV for the standard $N_{\rm eff} = 3.04$ scenario.
 \label{fig6} }
\end{figure}
Fig. \ref{fig6} shows the likelihood function marginalized over cosmological parameters for the case $N_{\rm eff}$ = 3.85, compared to the standard case $N_{\rm eff}$ = 3.04. The likelihood function is slightly larger at $m_\chi = \infty$ implying a better fit to the data, in agreement with the SPT results. The likelihood function also falls off more rapidly, resulting in a stronger mass bound $m_\chi > 8.6$ GeV compared to the bound $m_\chi > 7.6$ GeV obtained for $N_{\rm eff} = 3.04$. Ref. \cite{hooper_neutrinos, dark_rad, axino} have pointed out that the late time decay of a particle may mimic a neutrino-like degree of freedom. Axion dark matter models also prefer a larger value of $N_{\rm eff}$ \cite{sikivie_bec1, sikivie_bec2}.

\section{Dark matter halos and secondary CMB anisotropies}

Up to now, we have ignored the presence of dark matter halos. For weakly interacting dark matter, the earliest structures form at a redshift $z 
\sim 60$, and have masses $\gtrsim 10^{-6} M_\odot$ \cite{ghs1, ghs2}, although smaller minimum masses may be possible \cite{profumo_2006}. The minimum halo mass is set by the free streaming scale \cite{ghs1, ghs2} which in turn depends on the WIMP interaction cross section.

\subsection{Dark matter annihilation in halos}

Let us now look at the effect of dark matter halos on the evolution of the ionization fraction. The energy released per unit time and per unit volume $dE/dt dV$ is given by Eq. (\ref{wimp_ann}), but to account for halos, we must replace $\rho^2$ in Eq. (\ref{free}) by \cite{arvi1}:
\beq
\rho^2_{\chi, \rm halo} (z) = (1+z)^3  \, \int_{\rm M_{\rm min}} dM \, \frac{dn_{\rm halo}}{dM} (M,z) \left [ \int_0^{r_{\rm{200}}} dr \, 4 \pi r^2 \rho^2_{\rm h}(r) \right ] (M,z).
\label{halo}
\eeq
$n_{\rm halo}$ is the comoving number density of halos. The volume integral over the halo is a function of both halo mass $M$ and redshift $z$. We have assumed that halos are truncated at $r_{\rm 200}$, the radius at which the mean density enclosed equals 200 times the background density at the time of halo formation. $\rho_{\rm h}(r)$ is the halo density at a distance $r$ from the halo center. Eq. (\ref{halo}) ignores halo-halo interactions, as well as interactions between free dark matter and dark matter in halos. 

For dark matter halos with a generalized Navarro-Frenk-White (NFW) \cite{nfw} profile, we have
\begin{equation}
\rho_{\rm h}(r) = \frac{\rho_{\rm s}}   { (r/r_{\rm s})^ \alpha \left[1 + r/r_{\rm s}\right ]^\beta},
\label{nfw}
\end{equation}
and the volume integral over the halo takes the form \cite{arvi1}
\begin{eqnarray}
\int dr \, 4 \pi r^2 \, \rho^2_{\rm h}(r) &=& \frac{M \, \bar\rho}{3} \left( \frac{\Omega_{\rm dm}}{\Omega_{\rm m}} \right )^2 \, f(c) \nonumber \\
f(c) &=& \frac{c^3 \int_{\epsilon} ^{c} dx \, x^{2-2\alpha} \; (1+x)^{-2\beta}} { \left[ \int_{0}^{c} dx \, x^{2-\alpha} \; (1+x)^{-\beta} \right ]^2 }.
\end{eqnarray}
$M$ is the mass of the halo, and $c$ is the concentration parameter. $\bar \rho$ is the mean density of the halo which we set equal to 200 times the cosmological average matter density at the time of formation of the halo. $\epsilon$ is a dimensionless cutoff scale, required to make the luminosity finite for $\alpha > 1.5$. For the standard NFW profile ($\alpha = 1, \beta = 2, \epsilon = 0$), the parameter $f(c)$ is quite small. For example, for $c=10$, we find $f(10) \approx 150$. The value of $f(c)$ rises steeply when halos are more cuspy which could occur for e.g., due to adiabatic contraction. For $\alpha = 1.3$, we find $f(10) \approx 10^3$, and for $\alpha=1.5,\epsilon=10^{-5}$, we find $f(10) \approx 10^4$.

At moderately high redshifts $z>25$, inverse Compton scattering with CMB photons is the main mechanism by which high energy charged particles from WIMP annihilation lose energy to the CMB, producing a large number of medium energy ($\lesssim$ MeV) photons which ionize the gas. Here, we will assume that inverse Compton scattering is efficient, and the up-scattered CMB photons interact with gas atoms with a cross section $\sim \sigma_{\rm T}$. If the cross section may be thought of as independent of energy, the energy absorbed per gas atom, per unit time $\xi (z)$  is given by \cite{arvi3} (compare with Eq. (\ref{xi})):
\beq
\xi(z) = c \, \sigma_{\rm T} \int_\infty^z \frac{-dz'}{(1+z') H(z')} \, \left( \frac{1+z}{1+z'} \right )^3 \, \left( \frac{dE}{dt dV} \right ) (z')  \; e^{-\tau(z',z)}.
\label{energy_equation}
\eeq
The first term in the integrand comes from the relation $dz = -dt \, H(z) (1+z)$, while the next term accounts for the expansion of the Universe in the time it takes the ionizing radiation to reach redshift $z$ having been emitted at redshift $z'$. $\tau(z',z)$ is the optical depth from $z'$ to $z$. It can be shown that Eq. (\ref{energy_equation}) reduces to Eq. (\ref{xi}) in the limit $z' \approx z$, i.e. in the tight coupling limit, at high redshifts \cite{arvi3}.
\begin{figure}[!h]
\begin{center}
\scalebox{0.7}{\includegraphics{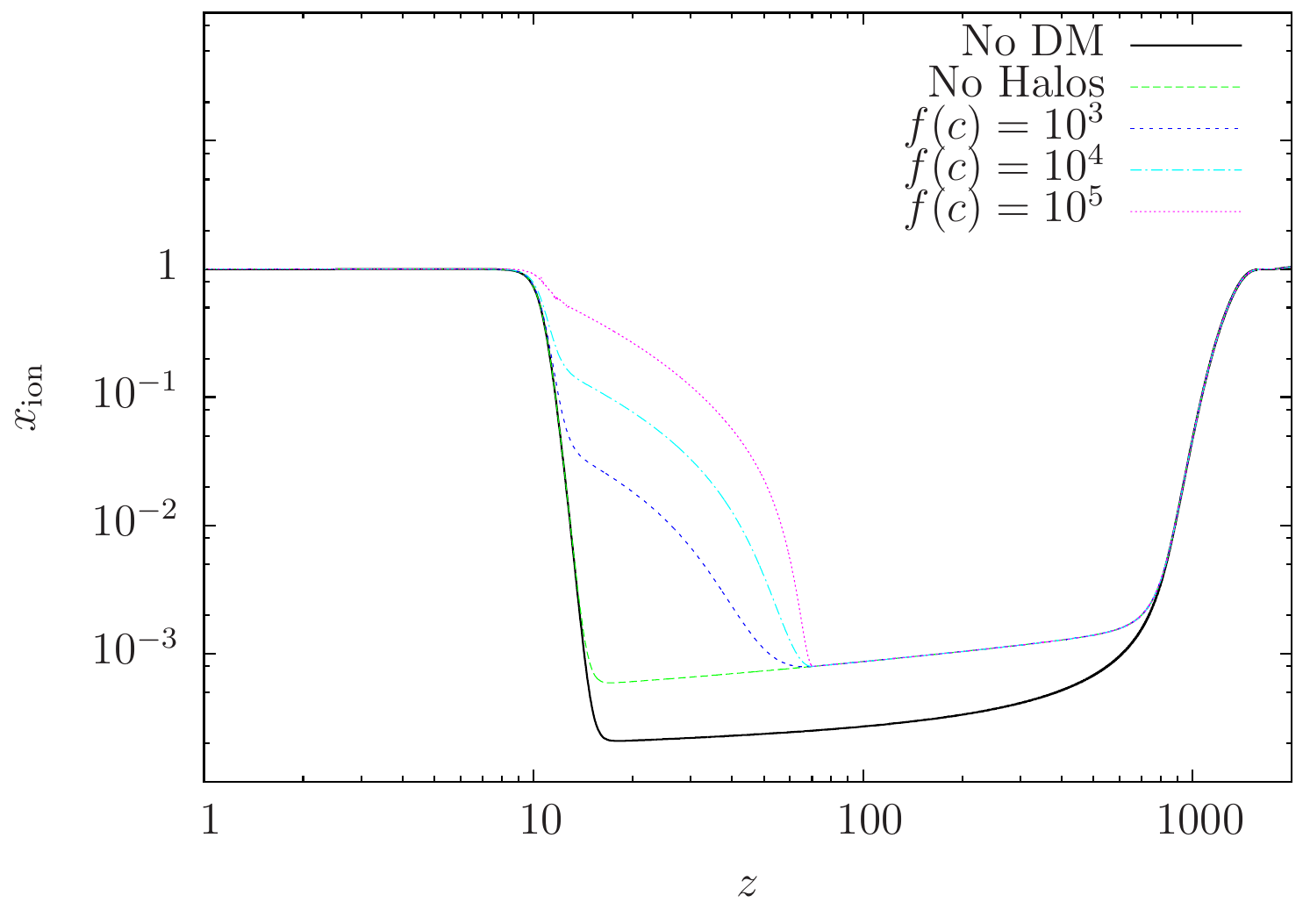}}
\end{center}
\caption{ 
Evolution of $x_{\rm ion}(z)$ in the presence of dark matter halos, for different values of $f(c)$. The solid (black) curve denotes the standard cosmological recombination model with no dark matter annihilation, while the dotted curves include dark matter annihilation. The green curve is plotted for free particle annihilation only, while the blue, cyan, and magenta curves account for dark matter halos. We set $M_{\rm min} = 10^{-9} M_\odot$, and use the Sheth-Tormen mass function.
 \label{fig7} }
\end{figure}
Fig. \ref{fig7} shows the evolution of the ionized fraction $x_{\rm ion}(z)$, for different cases. The solid black curve shows the case of no dark matter annihilation, while the green curve includes dark matter annihilation ($m_\chi = 10$ GeV), but ignores halos. The blue, cyan, and magenta curves include the effect of dark matter annihilation in halos. In all cases, one step reionization at $z_\ast \approx 10.5$ is assumed. For the rather extreme case of $f(c)=10^5$, the Universe is nearly completely ionized by $z=10$ by dark matter halos alone. Early ionization by halos may be tested by optical depth computations \cite{arvi1, belikov_hooper2009, cirelli_etal2009}. However, the WMAP constraints on the optical depth are largely dependent on the quality of the polarization power spectrum data which has a high signal-to-noise ratio only for small $l$. It is thus possible for very early ionization to be undetected by current data, although some constraints may be obtained from $TT$ and $TE$ data \cite{cmbpol}. For small values of $f(c)$, the primary CMB is not significantly affected by the inclusion of halos.

\subsection {The Ostriker-Vishniac effect}

The Ostriker-Vishniac (OV) / linear kinetic Sunyaev-Zeldovich effect is a second order effect caused by the scattering of CMB photons by electrons that have a peculiar motion. This effect was first studied by Zeldovich and Sunyaev \cite{zeldovich_sunyaev} and by Sunyaev and Zeldovich \cite{sunyaev_zeldovich}. It was studied in the context of large structure formation by Ostriker and Vishniac \cite{ostriker_vishniac}, and by Vishniac \cite{vishniac}. Here, we merely provide the formulae, following the work of Jaffe and Kamionkowski \cite{jaffe_kamion} (see also \cite{kheeganlee}).

The OV effect has been extensively studied in recent years as a probe of reionization. Partial reionization due to dark matter annihilation in halos differs from standard reionization in 2 aspects: (i) It is gradual and occurs at higher redshifts and (ii) it is more uniform as the mean free path of high energy photons/charged particles resulting from WIMP annihiation may be larger than the mean separation between halos. To our knowledge, the OV effect has not been applied to constrain partial reionization due to dark matter annihilation. We now examine whether the OV effect may be used to constrain dark matter annihilation in halos.

 The fractional temperature change induced by electrons with a bulk motion along the line of sight is \cite{vishniac}:
\beq
\frac{\Delta T}{T} = - \int dt \left( \hat n \cdot \vec v \right ) n_{\rm e} \sigma_{\rm T} e^{-\tau},
\eeq
where $\hat n$ is a unit vector denoting the line of sight, $\vec v$ is the peculiar velocity of the electrons, $n_{\rm e}$ is the number density of free electrons, $\sigma_{\rm T}$ is the Thomson cross section, and $\tau$ is the optical depth. In linear theory applicable at high redshifts, the peculiar velocity $\vec v$ may be simply expressed in terms of the matter overdensity. The power spectrum of temperature fluctuations will then be expressed in terms of the matter power spectrum, which is assumed to take the form given by Eisenstein and Hu \cite{eh}. The Ostriker-Vishniac power spectrum may then be written as:
\beq
C^{OV}_l =  \int_0^{\eta_0H_0} \frac{dx}{x^2} \; G^2(z) E^2(z) {\mathcal D}^2 (z) \, {\mathcal S}\left( \frac{l}{ \frac{c}{H_0} x} \right ).
\label{vish}
\eeq
$G, E, {\mathcal D}$, and $S$ are dimensionless quantities. $x = \eta_0 H_0 - \eta(z) H_0$ and
\beq
\eta(z) H_0  = \int_0^z \; \frac{dz'}{E(z')},
\eeq
and $E(z) = \sqrt{\Omega_{\rm m}(1+z)^3 + \Omega_\Lambda}$. $G$ and ${\mathcal D}$ are given by:
\bea
G(z) &=& E(z)   \frac{d\tau}{dz} e^{-\tau(z)} \n
{\mathcal D}(z) &=& \frac{D(z)}{D^2(0)} \, \frac {d D(z)}{dz}.
\eea
$D(z)$ is the growth function at redshift $z$. The function $S(k)$ takes the form
\beq
S(k) = \frac{1}{16 \pi^2} \left( \frac{H_0}{c} \right )^5 S_{\rm vish}(k),
\eeq
where $S_{\rm vish}(k)$ is the Vishniac power spectrum \cite{vishniac,jaffe_kamion}:
\beq
S_{\rm vish}(k) = k \int_0^\infty dy \int_{-1}^1 dx \, P(ky) P(k\sqrt{1+y^2-2xy}) \frac{ (1-x^2)(1-2xy)^2}{(1+y^2-2xy)^2}.
\eeq

\begin{figure}[!h]
\begin{center}
\scalebox{0.6}{\includegraphics{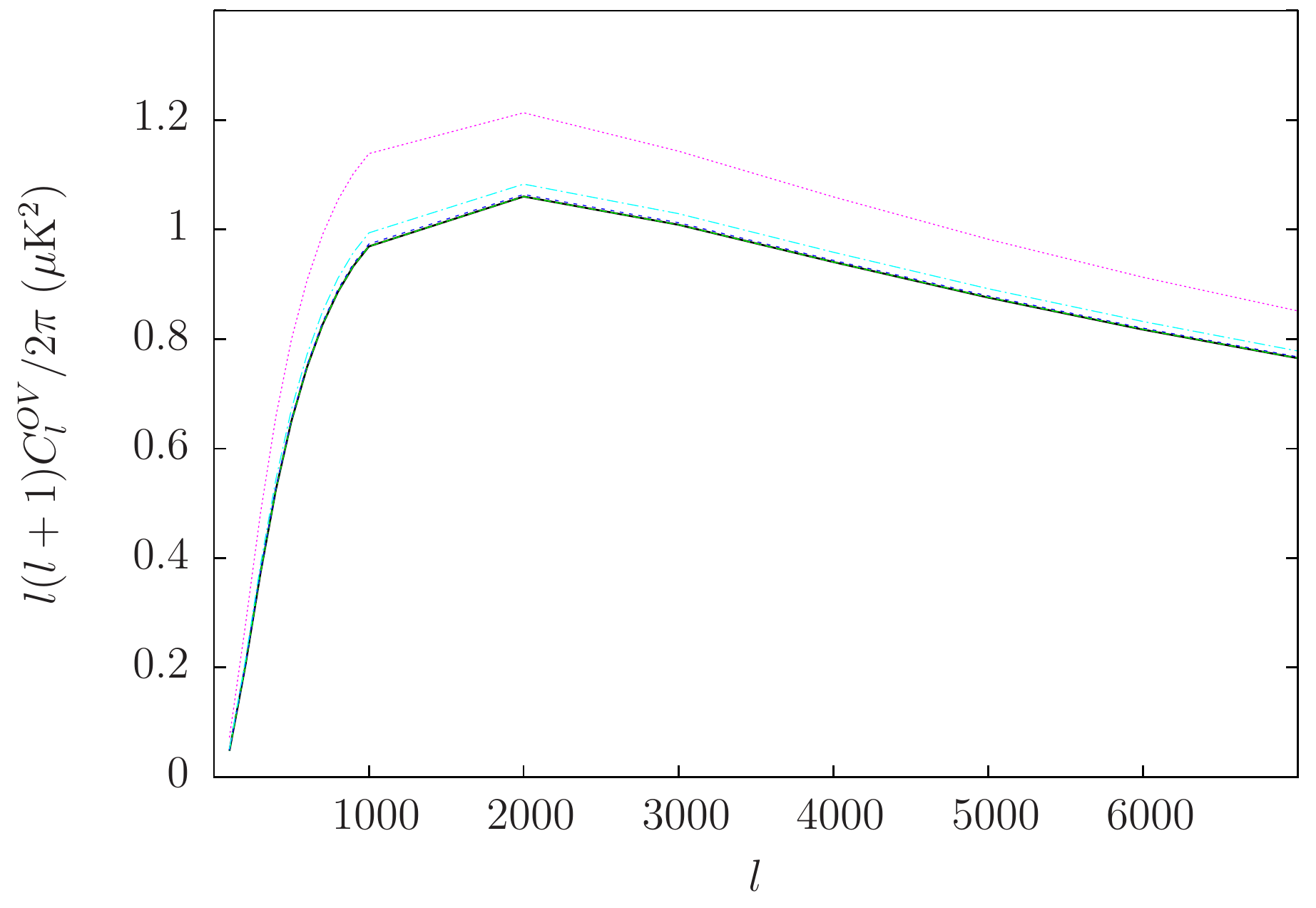}}
\end{center}
\caption{ 
The linear Ostriker-Vishniac power spectrum, for the different ionization histories shown in Fig. \ref{fig7}. Only the rather extreme case of $f(c)=10^5$ is distinguishable from the standard recombination history without dark matter annihilation.
 \label{fig8} }
\end{figure}

Let us apply Eq. (\ref{vish}) to dark matter reionization. Fig. \ref{fig8} shows the OV power spectrum for each of the dark matter models plotted in Fig. \ref{fig7}. Only the case of $f(c)=10^5$ is significantly different. Even in this case, the peak value differs from the standard recombination theory by only $\sim$ 20\%. One may hope to detect small changes in the OV power spectrum at frequencies that minimize the thermal SZ contribution. Even then, the contribution due to infrared sources needs to be modeled to high precision. The OV effect may be larger if the dark matter halos contain ionized baryons. In that case, one must take into account the bulk motion of the clusters themselves. The scenario is then similar to patchy reionization, and a full non-linear treatment is required. In the linear regime, we do not think that the OV effect is particularly useful in constraining dark matter annihilation due to halos, except perhaps for the most extreme cases. We however expect these extreme cases to be inconsistent with the primary CMB itself, thus limiting the usefulness of computing secondary anisotropies. Uncertainties in the reionization redshift $z_\ast$ also add to the difficulty in distinguishing different dark matter models.

\section{The large angle polarization power spectrum}

As discussed earlier, Thomson scattering of CMB photons by free electrons partially homogenizes the CMB temperature, leading to a damping in the CMB $TT$ power spectrum. Thomson scattering also causes additional polarization at scales $\sim$ the horizon at reionization, resulting in excess power at low multipoles in the $EE$ power spectrum.  Thus the large angle (small $l$) $EE$ power spectrum provides valuable information that is complementary to the information obtained from the $TT$ power spectrum.

\begin{figure}[!h]
\begin{center}
\scalebox{0.6}{\includegraphics{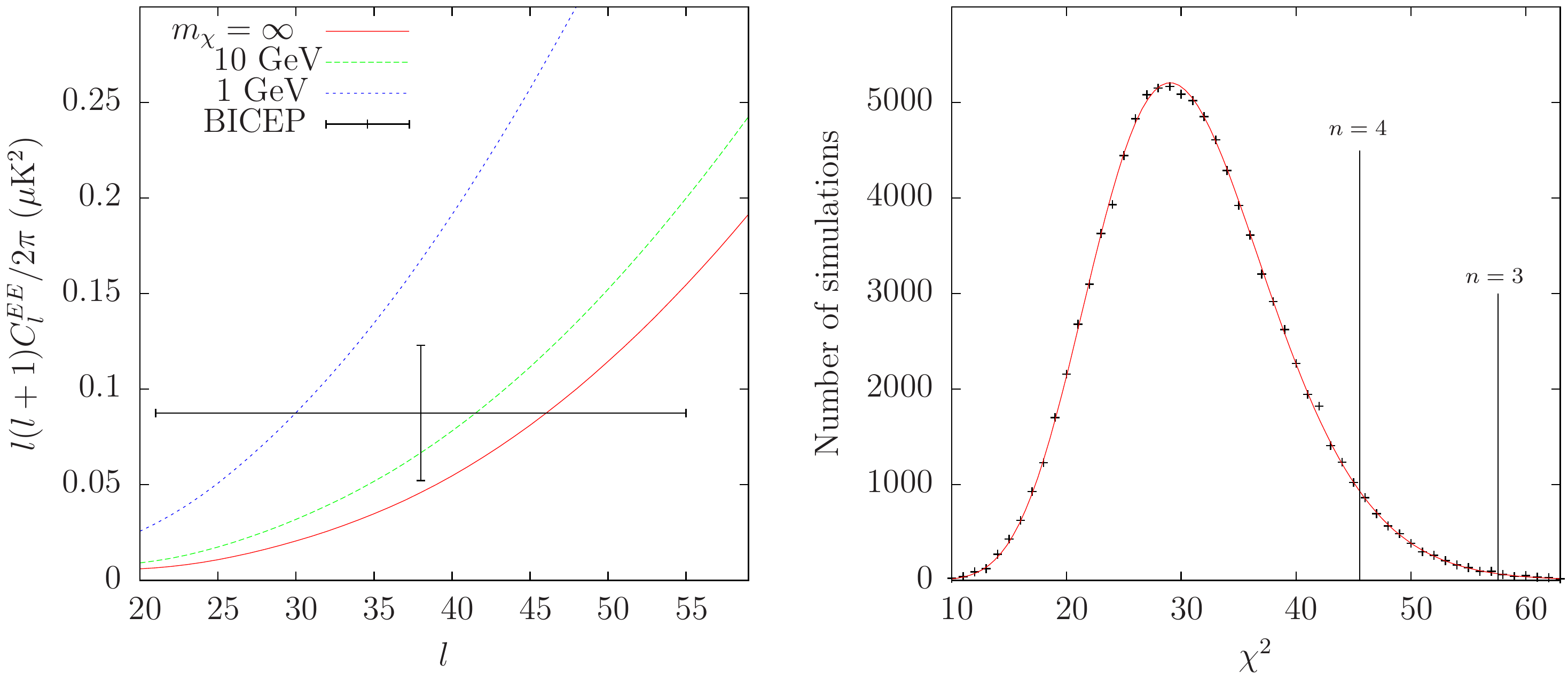}}
\end{center}
\caption{ 
(a) shows the large angle $EE$ power spectrum for $m_\chi = \infty$ (solid, red), as well as the cases with dark matter annihilation (dotted green and blue). Also shown is the first data point from the binned BICEP data release. (b) shows the result of $10^5$ Monte Carlo simulations for multipoles $20\leq l \leq 50$, assuming a $\Lambda$CDM model with no dark matter annihilation. The error bars are set to $n \, \times$ cosmic variance. The two vertical lines indicate the median value of $\chi^2_{\rm dm}$ for the $m_\chi$ = 10 GeV model, for $n=4$ and $n=3$. Only 4.4 (0.25)\% of the simulations (with no dark matter annihilation) result in a $\chi^2$ larger than the median value of $\chi^2_{\rm dm}$, for $n$ = 4 (3).
 \label{fig9} }
\end{figure}

Fig. \ref{fig9} shows the $EE$ power spectrum plotted for multipoles $20 \leq l \leq 50$ for the standard $\Lambda$CDM model, as well as models which include dark matter annihilation for $m_\chi$ = 10 GeV and $m_\chi$ = 1 GeV. Also shown is the first data point from the BICEP results (The WMAP experiment provides unbinned power spectra for $20 \leq l \leq 50$, but the data is prohibitively noisy). We have not included data for $l<20$ since the power spectrum on those scales is significantly affected by standard reionization by luminous sources. For $l>100$, the power spectrum is damped for the dark matter models, similar to the damping seen in the $TT$ and $TE$ power spectra. The current data is insufficient for the large angle $EE$ power spectrum to be a useful probe of dark matter annihilation. We therefore perform a number of Monte Carlo simulations to quantify the importance of the large angle polarization power spectrum.

We perform $10^5$ Monte Carlo simulations to predict the outcome of a real experiment. We restrict our discussion to light dark matter particles of mass $m_\chi \sim$ 10 GeV which are motivated by the results of direct detection experiments. We also assume that the correct theory is the standard $\Lambda$CDM model without WIMP dark matter annihilation,  and test the ability of future experiments to constrain dark matter with $m_\chi \sim$ 10 GeV. The assumed true parameters are determined by the cosmological model that provides the best fit to the $TT$ power spectrum data, and are given by $ \{ h=0.69, n_{\rm s} = 0.97, 10^9 A_{\rm s} = 2.245, \Omega_{\rm b}h^2 = 0.0225, \Omega_{\rm m}h^2 = 0.140  \}$. One-step reionization at $z_\ast=10.5$ is assumed.   For each value of $l$, the simulated $EE$ power spectrum $C^{EE}_l$ is a Gaussian distributed random number with a mean value given by the assumed theory, and a variance $\delta C^{EE}_l$ equal to $n \; \times$ cosmic variance (assuming full sky coverage):
\beq
\frac{\delta C^{EE}_l}{C^{EE}_l} = n \times \sqrt{\frac{2}{2l+1}}.
\eeq

We fit each Monte Carlo simulated data set from $l=20$ to $l=50$ with the assumed correct theory. In each case, we compute the value of $\chi^2$ (with 31 degrees of freedom). Fig. \ref{fig9}(b) shows the number of Monte Carlo simulations that result in a given value of $\chi^2$ within a bin of size $\Delta\chi^2 = 1$, along with the best fit $\chi^2$ distribution. We then fit each Monte Carlo simulation with the model that includes dark matter annihilation for $m_\chi$ = 10 GeV, assuming error bars of 4, and 3 $\times$ cosmic variance. The two solid lines show the median value of $\chi^2$, over $10^5$ simulations. For $n=4$, only 4.4\% of the Monte Carlo simulations result in a value of $\chi^2$ exceeding the median value of $\chi^2_{\rm dm}$ = 45.6 obtained for the $m_\chi$=10 GeV scenario. For $n=3$, only 0.25\% of the simulations result in a $\chi^2$ exceeding the median $\chi^2_{\rm dm}$ = 57.6. We therefore expect that a dark matter mass $m_\chi \sim$ 10 GeV may be excluded at the 95.6\% (99.7\%) confidence level provided the error bars are smaller than 4 (3) $\times$ cosmic variance. For comparison, the current WMAP unbinned $EE$ data has an error bar $\sim 47 \, \times$ cosmic variance at $l=40$.

It is important to note that the fit cannot be improved by varying $A_{\rm s}$, the parameter most degenerate with the effect of dark matter annihilation. This is because the value of $A_{\rm s}$ is fixed independently by the $TT$ power spectrum data. For the case of the standard theory without dark matter annihilation ($m_\chi \sim \infty$), the fit to the $TT$ power spectrum data yields $\chi^2_{\rm min}$ = 46.8/48 d.o.f. for $10^9 A_{\rm s}$ = 2.245, with the other parameters set to the values mentioned earlier. When the particle mass is reduced to $m_\chi = 10$ GeV \emph {with the value of $A_{\rm s}$ fixed}, one obtains $\chi^2$ = 254/48 d.o.f. which is conclusively ruled out by the data. However, the value of $A_{\rm s}$ is not fixed, and increasing $10^9 A_{\rm s}$ to 2.370 reduces the value of $\chi^2$ to 50.8/48 d.o.f., which while still disfavored by the data, is not excluded at high significance. Increasing the value of $A_{\rm s}$ to better fit the $TT$ data would \emph{worsen} the fit to the $EE$ data since a damping in the small angle $TT$ power spectrum is accompanied by a boost in the large angle $EE$ power spectrum. With the new value of $A_{\rm s}$, the median $\chi^2$ for the $n$ = 4 (3) cases increases from 45.6 (57.6) to 51.3 (67.8) with 31 d.o.f. 

Let us now consider the Planck experiment. The error bar at multipole $l$ is given by \cite{knox_1995, jungman_1996}:
\beq
\frac{\delta C^{EE}_l}{C^{EE}_l} = \sqrt{\frac{2}{2l+1}} \, \frac{1}{f^{1/2}_{\rm sky}} \; \left [ 1 + \frac{ (f_{\rm sky} w)^{-1}}    {C^{EE}_l} e^{l(l+1)\sigma^2_{\rm b}} \right ].
\eeq
$w^{-1} = \sigma^2_{\rm pix} \, \theta^2_{\rm fwhm}$ is the inverse weight per solid angle. $\sigma_{\rm pix}$ is the pixel noise, $\theta_{\rm fwhm}$ is the beam full width at half maximum, $f_{\rm sky}$ is the fraction of sky covered, and $\sigma_{\rm b} = \theta_{\rm fwhm}/\sqrt{8 \ln 2}$. The number of pixels $\approx 4\pi / \theta^2_{\rm fwhm}$.  The inverse weight per solid angle is then given by $w^{-1} = 4 \pi (\Delta T)^2/ ( t \times n_{\rm bol})$, where $\Delta T$ is the noise equivalent temperature per bolometer,  $t$ is the observation time, and $n_{\rm bol}$ is the number of bolometers for the given frequency channel.  For the 143 GHz polarization sensitive channel of Planck (143P), $\Delta T$ = 82 $\mu$K$\sqrt{{\rm s}}$, $n_{\rm bol}$ = 8, and $\theta_{\rm fwhm} = 7'$  \cite{planck_hfi}, giving us $w^{-1} = 2.7 (1.1) \times 10^{-4}  \mu$K$^2$ for  $t$ = 15 months (3 years). Assuming $f_{\rm sky} \approx 0.65$ \cite{planck}, we find that the Planck mission can exclude $m_\chi$ = 10 GeV with $EE$ power spectrum data from $20<l<50$ at $\lesssim 2\sigma$ with 15 months observation time and at $> 3 \sigma$ significance with 3 years observation time. The combined $TT + TE + EE$ data set from Planck will provide even better constraints.

\section{Conclusions}
In this article, we have examined how current CMB data can set limits on WIMP dark matter annihilation, for the simple models in which the WIMP is all of the dark matter, with $s-$wave dominated annihilation at the thermal rate $\acs = 3 \times 10^{-26}$ cm$^3$/s. Unless the dark matter annihilates primarily into neutrinos, one may probe dark matter masses $m_\chi < 10$ GeV using current CMB data. Future data from the Planck mission is expected to substantially improve this bound.

In Section II, we discussed the physics of dark matter annihilation, and computed the electromagnetic fraction $f_{\rm em}$ for the annihilation channels $\chi \chi \rightarrow b \bar b, c \bar c, \tau^+ \tau^-$. We studied dark matter annihilation at high redshifts, and obtained an expression for the energy absorbed per gas atom, and the ionization and temperature evolution with redshift. 

In Section III, we discussed how the CMB power spectra are modified by dark matter annihilation at high redshifts. We performed a maximum likelihood analysis using the publicly available CMB Boltzmann code CLASS, and CMB data from the WMAP, SPT, BICEP, and QUaD experiments. We obtained the likelihood as a function of WIMP mass $m_\chi$ by marginalizing over the cosmological parameters $A_{\rm s}, n_{\rm s}, h, \Omega_{\rm b} h^2, \Omega_{\rm m} h^2$. For the $\chi \chi \rightarrow b \bar b$ channel, we found that WIMP masses $m_\chi < 7.6$ GeV are excluded at the 95\% confidence level, for the simplest dark matter models. We thus find that direct detection experiments that prefer a mass $m_\chi \sim$ 10 GeV are consistent with CMB data. We then investigated whether constraints on $\sigma_8$ from galaxy cluster observations may improve the bound on $m_\chi$ by restricting the range over which $A_{\rm s}$ may vary. Unfortunately, the large degeneracy between $\sigma_8$ and $\Omega_{\rm m}$ does not allow a precise determination of $\sigma_8$, and hence the CMB bound on $m_\chi$ is only marginally improved. The bound on $m_\chi$ is strengthened slightly if $N_{\rm eff}$ is accurately determined to be larger than the standard model value of $N_{\rm eff} = 3.04$. For $N_{\rm eff} = 3.85$, we obtain $m_\chi > 8.6$ GeV (95\% confidence, $b\bar b$ channel) which is better than the bound $m_\chi > 7.6$ GeV obtained for $N_{\rm eff} = 3.04$.

In Section IV, we studied the effect of dark matter halos on the reionization history of the Universe. We showed that for optimistic halo parameters, it is possible for dark matter halos to substantially reionize the Universe. We obtained an expression for the energy absorbed by gas atoms at a redshift $z$ due to particle annihilation at redshift $z'$, and solved for the evolution of the ionization fraction as a function of redshift. We then discussed the Ostriker-Vishniac (OV) effect as a possible probe of dark matter annihilation at intermediate redshifts $10<z<60$. We computed the OV power spectrum for different ionization histories, but found the effect to be small except in the most extreme cases.

In Section V, we examined the importance of the large angle polarization power spectrum as a probe of dark matter annihilation that provides information complementary to what is obtained from the temperature power spectrum. Since current experiments do not provide high quality polarization data for $20<l<50$, we performed a number of Monte Carlo simulations and showed that the $EE$ power spectrum can help constrain dark matter properties. We expect that $m_\chi = 10$ GeV may be excluded at the 2 (3) $\sigma$ level using $EE$ data from $20<l<50$ provided the error bars are smaller than 4 (3) $\times$ the cosmic variance value. We expect the upcoming Planck results to provide significant improvements on the minimum allowed WIMP dark matter mass. The improved mass bound will be relevant to direct detection experiments.

\acknowledgments{ I am grateful to Nick Battaglia, Nishikanta Khandai, Hy Trac, Andrew Hearin, and Arthur Kosowsky for many helpful discussions. I thank Sean Bryan for making me aware of the BICEP data results. I acknowledge financial support from the Bruce and Astrid McWilliams Center for Cosmology. }

\end{document}